\documentclass[]{raa}            
\usepackage{graphicx,times}
\usepackage{natbib}

\providecommand{\bm}[1]{\mbox{\boldmath$#1$\unboldmath}}

\begin{document}

   \title{Aggregate dust model to study the polarization properties of comet C/1996 B2 Hyakutake
}

 \volnopage{ {\bf 2009} Vol.\ {\bf 9} No. {\bf XX}, 000--000}
   \setcounter{page}{1}

   \author{H. S. Das
   \and A. Suklabaidya
 \and S. Datta Majumder
      \and A. K. Sen
   }

   \institute{Department of Physics, Assam University, Silchar
              788011, India; {\it hsdas@iucaa.ernet.in}\\
\vs \no
   {\small Received [year] [month] [day]; accepted [year] [month] [day] }
}

\abstract{ In our present study, the observed linear polarization data of comet Hyakutake are
studied at wavelengths $\lambda=0.365\mu m$,  $\lambda=0.485\mu m$ and 0.684$\mu m$ through
simulations using Ballistic Particle-Cluster Aggregate and Ballistic Cluster-Cluster Aggregate
aggregates of 128 spherical monomers.  We first investigated that the size parameter of the
monomer, $x$ $\sim$ 1.56 -- 1.70, turned out to be most
suitable which provides the best  fits to the observed dust scattering properties at
three wavelengths $\lambda = 0.365$$\mu m$, 0.485$\mu m$ and 0.684$\mu m$. Thus  the effective
radius of the aggregate (r) lies in the range $0.45 \mu m \le r \le 0.49 \mu m$
at $\lambda = 0.365$$\mu m$; $ 0.60 \mu m \le r \le 0.66 \mu m$ at $\lambda = 0.485$$\mu m$
and $0.88 \mu m \le r \le 0.94 \mu m$ at $\lambda = 0.684$$\mu m$. Now using
superposition \textsc{t-matrix} code and the power-law size distribution, $n(r) \sim r^{-3}$, the best-fitting
values of complex refractive indices are calculated which can best fit the
observed polarization data at the above three wavelengths.  The best-fitting complex refractive indices $(n,k)$
 are found to be (1.745, 0.095) at $\lambda = 0.365 $ $\mu m$, (1.743, 0.100) at $\lambda = 0.485 $ $\mu m$  and
(1.695, 0.100) at $\lambda = 0.684$ $\mu m$.  The refractive indices coming out from the present analysis correspond to
mixture of both silicates and organics, which are in good agreement with the \textit{in situ}
measurement of comets by different spacecraft.
}

   \authorrunning{H. S. Das,  A. Suklabaidya, S. Datta Majumder \& A. K. Sen}            
   \titlerunning{Polarization properties of comet C/1996 B2 Hyakutake }  
   \maketitle


%
%
\section{Introduction}           
\label{sect:intro}

The study of cometary polarization,
over various scattering angles and wavelengths, gives valuable
information about the nature of cometary dust. The  numerical and experimental simulations  of
polarization data gives  information about the physical properties
of the cometary dust, which include size distribution, shape and
complex refractive indices.  Several investigators (Kikuchi et al. 1987;
Lamy et al. 1987; Sen et al. 1991a, 1991b; Chernova et al. 1993; Xing \& Hanner 1997;
Petrova et al. 2004; Kimura et al. 2006; Das et
al. 2004; Kolokolova et al. 2007, Bertini et al. 2007 etc.) have
studied linear and circular polarization measurements of many
comets. These studies help us to understand the dust grain
nature of comets.

 Comet Hyakutake (C/1996 B2) was the brightest comet appeared in the sky in the year 1996.  Its passage near the Earth was one of the closest cometary approaches of the previous 200 years which passed within 0.1 AU of the Earth in March 1996.  Comet Hyakutake  was bright enough to make high precision polarimetric observations during pre-perihelion phase. Observations for the linear polarization of comet Hyakutake were carried out at three different wavelengths: 0.365$\mu m$ 0.485$\mu m$ and 0.684$\mu m$ by different investigators (Joshi et al. 1997; Kiselev \& Velichko 1998 and Manset \& Bastien 2000).

Greenberg and Hage (1990) first suggested that cometary particles are not spherical and porous.
They originally proposed the presence of large
numbers of {\it porous} grains in the coma of comets to explain
the spectral emission at 3.4$\mu m$ and 9.7$\mu m$. Dollfus (1989)
discussed the results of laboratory experiments by microwave
simulation and laser scattering on various complex shapes with
different porosities. The results of {\it in situ} measurements
carried out on the Giotto spacecraft at Comet Halley (Fulle et al.
2000) and the analysis of the infrared spectra of Comet Hale-Bopp
(Moreno et al. 2003) also agree with the model of aggregates. It
is clear from recent modeling of optical (Xing \& Hanner 1997;
Kimura 2001; Kimura et al. 2006; Petrova et al.
2004; Tishkovets et al. 2004;  Lasue et al. 2006; Kolokolova et al. 2007;
Bertini et al. 2007;  Levasseur-Regourd et al. 2007,  2008; Das et al. 2008a; 2008b etc.),
thermal-infrared observations (Lisse et al. 1998; Harker et al.
2002), laboratory studies (Wurm \& Blum 1998; Gustafson \&
Kolokolova 1999; Hadamcik et al. 2002 etc.), and especially from
the `Stardust' returned samples ( H\"{o}rz et al.  2006), that
cometary dust consists of irregular, mostly aggregated particles.

 Das \& Sen (2006) studied the non-spherical dust grain characteristics of Comet Levy 1990XX
using the T-matrix theory. They found that compact prolate
grains as compared to spherical grains can better explain the
observed linear polarization data.  Recently, Das et al. (2008a) have again analyzed
the observed polarization data of Comet Levy 1990XX and successfully reproduced
the polarization curve through simulations using aggregate dust
model, where the fit was still better. It has been found from their analysis that aggregate
particles can produce a still better fit to the observed data as compared to compact prolate grains.
Recently, using aggregate dust model, Das et al. (2008b) successfully
explained the polarization characteristics of comet Hale-Bopp at
$\lambda = 0.485$ $\mu m$ and 0.684 $\mu m$. Lasue et al. (2009) have explained successfully the
polarization properties of comet Hale-Bopp and Halley by using a model of light scattering through
a size distribution of aggregates (spherical or spheroidal) mixed with single spheroidal particles.

In the present work, the aggregate dust model is proposed to study
the observed polarization data of Comet Hyakutake at $\lambda$ = 0.365$\mu m$,
0.485$\mu m$ and 0.684$\mu m$.


\section{Aggregate model of cometary dust:}
\label{sect:Agg}

The aggregates are built by using ballistic aggregation procedure. Two types of aggregates
are considered here- BPCA (Ballistic Particle-Cluster Aggregate) and BCCA (Ballistic Cluster-Cluster Aggregate).
In actual case, the BPCA clusters are more compact
than BCCA clusters (Mukai et al. 1992).  A systematic explanation on dust aggregate model is
already discussed in our previous work (Das et al. 2008a).
 Laboratory diagnosis of particle
coagulation in the solar nebula suggests that the particles grow under BCCA process.
It is also found that the morphology of dust particles does not play a major role
in determining the shape of polarization (Kimura 2001; Kimura et al. 2003, 2006;
Kolokolova et al. 2006; Lasue \& Levasseur-Regourd 2006; Bertini et al. 2007;
Das et al. 2008a, 2008b). The size of the individual monomer in a cluster plays an important
role in scattering calculations. These have been confirmed by the
results of previous work on dust aggregate model (Kimura et al.
2003; Kimura et al. 2006; Petrova et al. 2004; Hadamcik et al. 2006; Bertini et
al. 2007) and also from our previous work (Das et al. 2008a).

\section{Composition}
The \textit{in situ} observations of comets, laboratory analysis of samples of Interplanetary Dust Particles
(IDP) and remote infrared spectroscopic study of comets give useful information
about the composition of cometary dust.  The \textit{in situ}
measurement of impact-ionization mass spectra of Comet Halley's
dust, has suggested that the dust consists of magnesium-rich
silicates, carbonaceous materials, and iron-bearing sulfides
( Kissel et al. 1986; Jessberger et al. 1988 and Jessberger 1999).
Actually, the first evidence for carbonaceous material in comets  comes from
the study of Vega spacecraft data by Kissel et al. (1986). These materials are
also known to be the major constituents of IDPs (Brownlee et al.
1980). The studies of comets and IDPs have
shown the presence of amorphous and crystalline silicate minerals
(e.g.  forsterite, enstatite) and organic materials (Hanner \&
Bradley 2004). Laboratory studies have  shown that majority of the
collected IDPs fall into  the spectral classes defined by their
10 $\mu m$  feature profil. These observed profiles indicate the presence
of  pyroxene, olivine and layer lattice silicates. This is in good agreement with results
obtained from Giotto and Vega mass spectrometer observations of
Comet Halley (Lamy et al. 1987). The infrared (IR) measurement of
comets has also provided important information on the silicate
compositions in cometary dust. The spectroscopic studies of
silicates have shown the predominance of both crystalline and
amorphous silicates consisting of pyroxene or olivine grains
(Wooden et al. 1999; Hayward et al. 2000, Bockel\'{e}e - Morvan et
al. 2002 etc.). Mg-rich crystals are also found within IDPs and
are predicted by comparing the IR spectral features of Comet
Hale-Bopp with synthetic spectra obtained from laboratory studies
(Hanner  1999; Wooden et al. 1999, 2000).  `Stardust' samples have
also confirmed a variety of olivine and pyroxene silicates in
Comet 81P/Wild 2 (Zolensky et al. 2006).

Levasseur-Regourd et al. (1996) studied a polarimetric
data base of several comets and from the nature of phase
angle ($\alpha$) dependence, they concluded that there is a clear evidence
for at least two classes of comets according to the values of polarization  at
$\alpha \approx 80^0 - 100^0$: comets with a high maximum in polarization, of about 25\%
for one group  and smaller than 15\% for the other
group. The two classes of comets are distinct only for $\alpha > 35^0$.
 It has been also observed that there is a very good correlation between the existence of a high maximum in polarization and a strong silicate emission feature (Levasseur-Regourd 1999). The observed polarization data of comet Hyakutake  showed a high maximum in polarization. The polarization at a given phase angle larger than 30$^0$ most often increases linearly with increasing wavelength in the visible domain and this increase being steeper for larger phase angles (Levasseur-Regourd \& Hadamcik 2003). Recent studies have provided useful information about the
two groups of polarimetrically different comets
(Kiselev et al., 2001; Kiselev et al., 2004; Jewitt 2004; Jockers et al., 2005).

It has been already found that the silicate composition can best reproduce the observed polarization data
of Comet Levy 1990XX and Comet Hale-Bopp (Das et al. 2008a,b).

\section{Numerical simulations:}
\label{sect:num}

The scattering calculations for BCCA \& BPCA particles have been done  by the Superposition T-matrix code, which gives
rigorous solutions for ensembles of spheres (Mackowski \&
Mishchenko 1996). The observed linear polarization data of comet Hyakutake at  $\lambda$ = $0.365\mu m$, $0.485\mu m$ \& $0.684\mu m$ are taken from Joshi et al. (1997), Kiselev \& Velichko (1998), Manset \& Bastien (2000).

 The linear polarization is given by

\begin{equation}
    P(\theta) = - \frac{S_{21}}{S_{11}}
\end{equation}

 For modeling comet Hyakutake, we will use a power-law size distribution, $n(r) = dn/dr \sim r^{-3}$. For a particular type of aggregate with fixed N, the size distribution
is just $dn/da_m \sim a_m^{-3}$. Thus the averaged polarization is (Shen et al. 2009):

\begin{equation}
    \bar{P} = \frac{\int_{a_{min}}^{a_{max}} p(a_m, \theta)\, S_{11}(a_m, \theta)\, n(a_m)\, da_m}{\int_{a_{min}}^{a_{max}}S_{11}(a_m, \theta)\, n(a_m) \, da_m}
\end{equation}
 where $a_{min}$ and $a_{max}$ are the minimum  and maximum values of the monomer size in our size distribution.

 The radius of an aggregate particle can be described by the radius
of a sphere of equal volume given by $r = a_m N^{1/3}$, where N is the number of monomers in the aggregate. In the present work, N =128 is taken. The size parameter of the monomer is given by $x = \frac{2\pi a_m}{\lambda}$.
We first investigated that $x$ $\sim$ 1.56 -- 1.70 turned out to be most suitable which may provide the best qualitative fits to the observed dust scattering properties at three wavelengths $\lambda = 0.365$$\mu m$, 0.485$\mu m$ and 0.684$\mu m$. This correspond to  $0.090\mu m \le a_m \le 0.098 \mu m$ at $\lambda = 0.365$$\mu m$; $0.120\mu m \le a_m \le 0.131 \mu m$ at $\lambda = 0.485$$\mu m$ and
$0.174\mu m \le a_m \le 0.186 \mu m$ at $\lambda = 0.684$$\mu m$. Thus  the effective radius of the aggregate (r) lies in the range $0.45\mu m \le r \le 0.49 \mu m$ at $\lambda = 0.365$$\mu m$; $0.60\mu m \le r \le 0.66 \mu m$ at $\lambda = 0.485$$\mu m$ and $0.88\mu m \le r \le 0.94 \mu m$ at $\lambda = 0.684$$\mu m$.

We start calculations considering the refractive indices for amorphous pyroxene and amorphous olivine at $\lambda$=$0.365\mu m$, $0.485\mu m$ \& $0.684\mu m$. The refractive indices of the materials are calculated by linearly interpolating the data obtained from laboratory studies (Dorschner et al. 1995).  Olivines and pyroxenes
are described by Mg$_{2y}$Fe$_{2-2y}$SiO$_4$, with $y$ = 0.4,
0.5 and  Mg$_{y}$Fe$_{1-y}$SiO$_3$, with $y$ = 1.00, 0.95, 0.8, 0.7, 0.6, 0.5 and 0.4.


It has been already investigated that the choice of the above values of
refractive indices can't match the observed polarization data of
Comet Levy 1990XX and Hale-Bopp (Das et al. 2008a,b). The same set
of refractive indices is now chosen to fit the observed
polarization data of Comet Hyakutake. The calculations have been done
for BCCA aggregates. But no such good fit has been observed using
the above value of refractive indices. Next, the calculation has been
repeated for  carbonaceous materials,
but none of them could match the observed data well.

We now use $\chi ^2$ minimization technique to evaluate the best-fitting
values of $(n,k)$, which can fit to the observed
polarization data. We have already used this minimization
technique to fit the observed linear polarization data of Comet
Levy 1990XX at $\lambda = 0.485\mu m$ and Comet Hale-Bopp at
$\lambda = 0.485\mu m$ and $0.684\mu m$ (Das et al. 2008a,b), with aggregate
models of dust.

The error in the fitting procedure can be defined as

\begin{equation}
\chi_{\textrm{pol}}^2 = \sum_{i=1}^J  \left|
\begin{array}{c}
\frac{\textrm{P}_{obs} (\theta_i,\lambda) -
\textrm{P}_{model}(\theta_i,\lambda)} {
 \textrm{E}_p (\theta_i,\lambda)}
\end{array}
\right|^2
\end{equation}

Here, $\textrm{P}_{obs} (\theta_i,\lambda)$ is the degree of
linear polarization observed at scattering angle $\theta_i$ (i =
1,2,....,J) and wavelength $\lambda$,
$\textrm{P}_{model} (\theta_i,\lambda)$ is the polarization values
obtained from model calculations and $\textrm{E}_p
(\theta_i,\lambda)$ is the error in the observed polarization at
scattering angle $\theta_i$ and wavelength ($\lambda$).

We now introduce a quantity $ \chi ^2 = \chi_{\textrm{pol}}^2 /J
$, where J is the number of data points. The values of $(n,k)$ are
varied over a large range simultaneously with $a_m$ and we find for a particular value of
$(n,k)$, $\chi ^2$ becomes minimum. This particular value of
$(n,k)$ is our best fitted $(n,k)$ value and the corresponding
minimum value of $\chi ^2$ is denoted as $\chi_{\textrm{min}}^2$.
It is also observed that this technique of minimization of $\chi
^2$ is quite unique. The value of $\chi_{\textrm{min}}^2$ gives
the confidence level on our best fit values of $(n,k)$ and also in
the overall fitting procedure.

We need to fine-tune the free parameters $(n,k)$ in the model to
make the best fit to the observed linear polarization data of Comet
Hyakutake. The real part of the refractive index is increased from $n$ = 1.4 to 2.0
with 0.001 steps, while the imaginary part of the refractive index
is increased from $k$ = 0.001 to 1.0 with 0.001 steps. The same
range has been already used for Comet Levy 1990XX and Comet
Hale-Bopp (Das et al. 2008a,b).

 Now we analyze the observed data of comet Hyakutake at 0.365$\mu m$. We calculate $\bar{P}$ averaged over
the size distribution $n(r) \sim r^{-3}$ with $r_{min} = 0.45$$\mu m$  and $r_{max} = 0.49 \mu m$
 ($a_{min} = 0.090 \mu m$, $a_{max} = 0.098 \mu m$). The best fitting
refractive index at 0.365$\mu$m is found to be n = 1.745, k = 0.095.
The simulated  polarization curve at 0.365$\mu m$ is shown in \textbf{Fig. 1}.

 We now extend our calculation further to fit the observed polarization data
at $\lambda$ = 0.485$\mu m$ and 0.684 $\mu m$. Here we also calculate  $\bar{P}$ averaged
over the size distribution $n(r) \sim r^{-3}$ with $r_{min} = 0.60\mu m$ and $r_{max} = 0.66\mu m$  ($a_{min} = 0.120 \mu m$, $a_{max} = 0.131 \mu m$) at $\lambda$ = 0.485$\mu m$
and $r_{min} = 0.88\mu m$ and $r_{max} = 0.94\mu m$  ($a_{min} = 0.174 \mu m$,  $a_{max} = 0.186 \mu m$)
at $\lambda$ = 0.684$\mu m$. The best fitting refractive indices are obtained from the
present analysis are found to be (1.743, 0.100) at $\lambda$ = 0.485$\mu m$ and (1.695, 0.100) $\lambda$ = 0.684$\mu m$. The simulated polarization curve
for comet Hyakutake at $\lambda$ = 0.485$\mu m$ and
0.684 $\mu m$  for BCCA aggregates are shown in \textbf{Fig. 2} and \textbf{Fig. 3}.

  \begin{figure*}
  \centering
  \includegraphics[scale=1, angle=0]{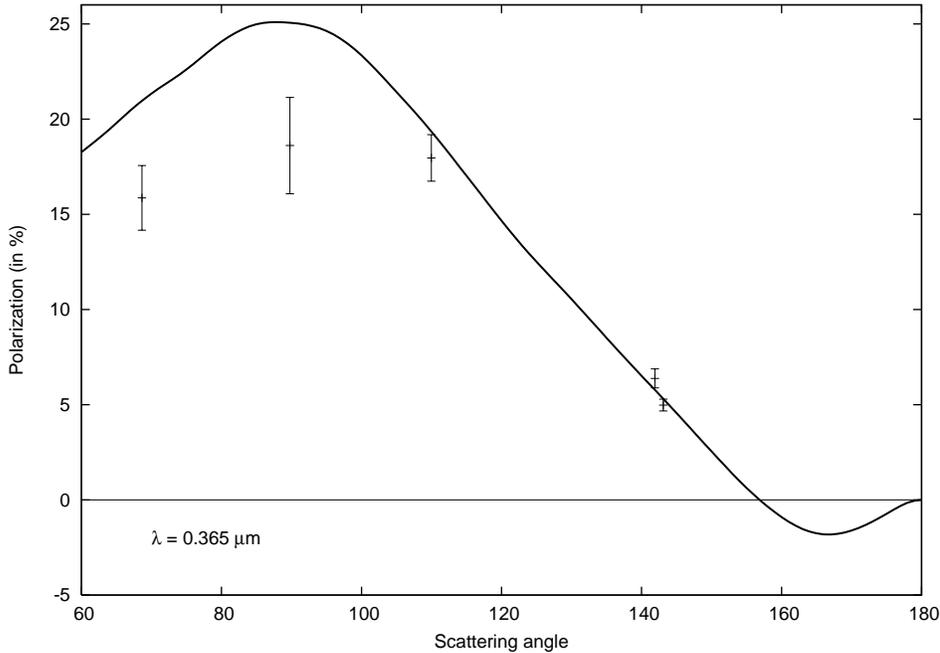}
   \caption{Polarization values as observed at wavelength $\lambda$ = 0.365$\mu m$ for
comet Hyakutake by Joshi et al. (1997) and Kiselev \& Velichko (1998). The solid  curve represents
 the best-fitting polarization curve  obtained  for BCCA particles
 with 128 monomers  for a size distribution $n(r) \sim r^{-3}$ for $0.45\mu m \le r \le 0.49 \mu m$ at
 $\lambda = 0.365$ $\mu m$. Here $n$ = 1.745, $k$ = 0.095.}\label{figexample}
    \end{figure*}

  \begin{figure*}
  \centering
  \includegraphics[scale=1, angle=0]{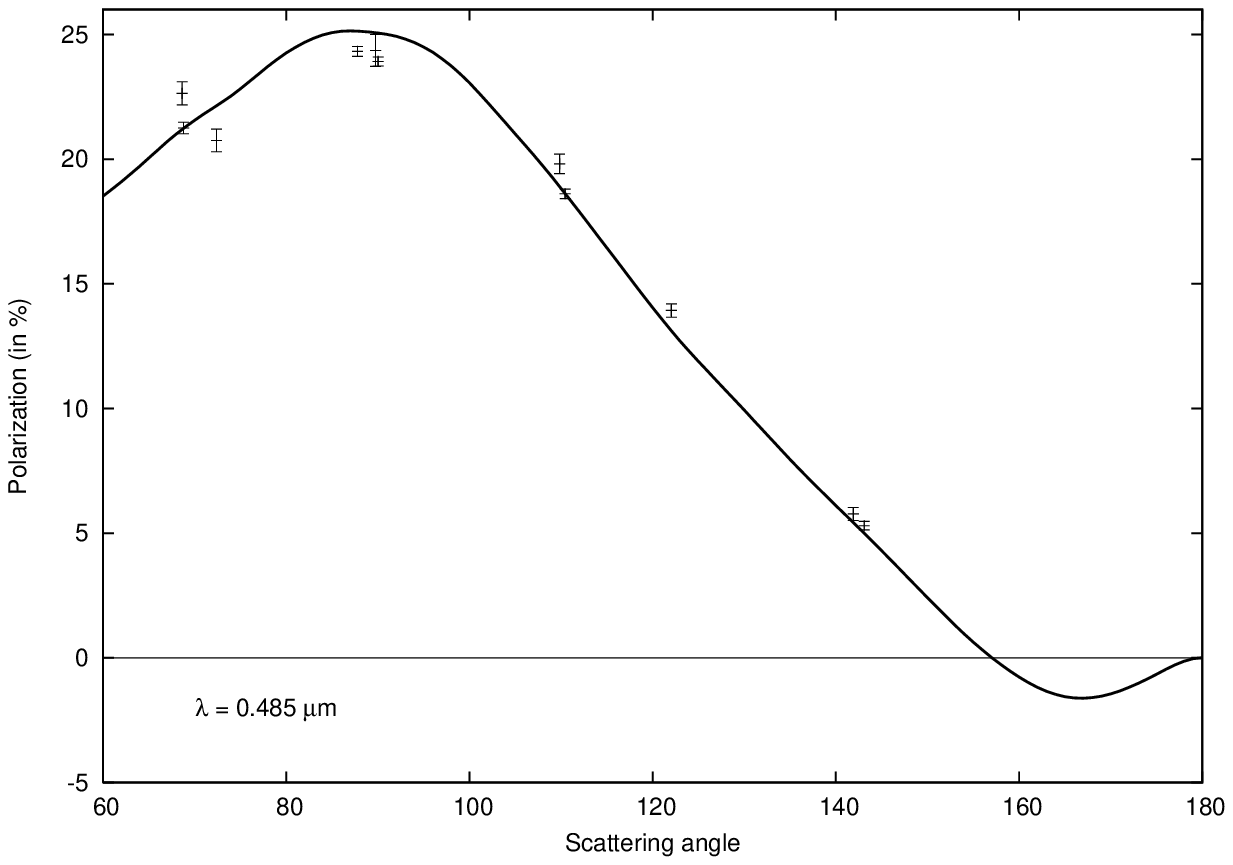}
   \caption{Polarization values as observed at wavelength $\lambda$ = 0.485$\mu m$ for
comet Hyakutake by Joshi et al. (1997) and Kiselev \& Velichko (1998). The solid  curve represents
 the best-fitting polarization curve
 obtained  for BCCA particles
 with 128 monomers  for a size distribution $n(r) \sim r^{-3}$ for $0.60\mu m \le r \le 0.66 \mu m$ at
 $\lambda = 0.485$ $\mu m$. Here $n$ = 1.743, $k$ = 0.100.}\label{figexample}
    \end{figure*}

  \begin{figure*}
  \centering
  \includegraphics[scale=1, angle=0]{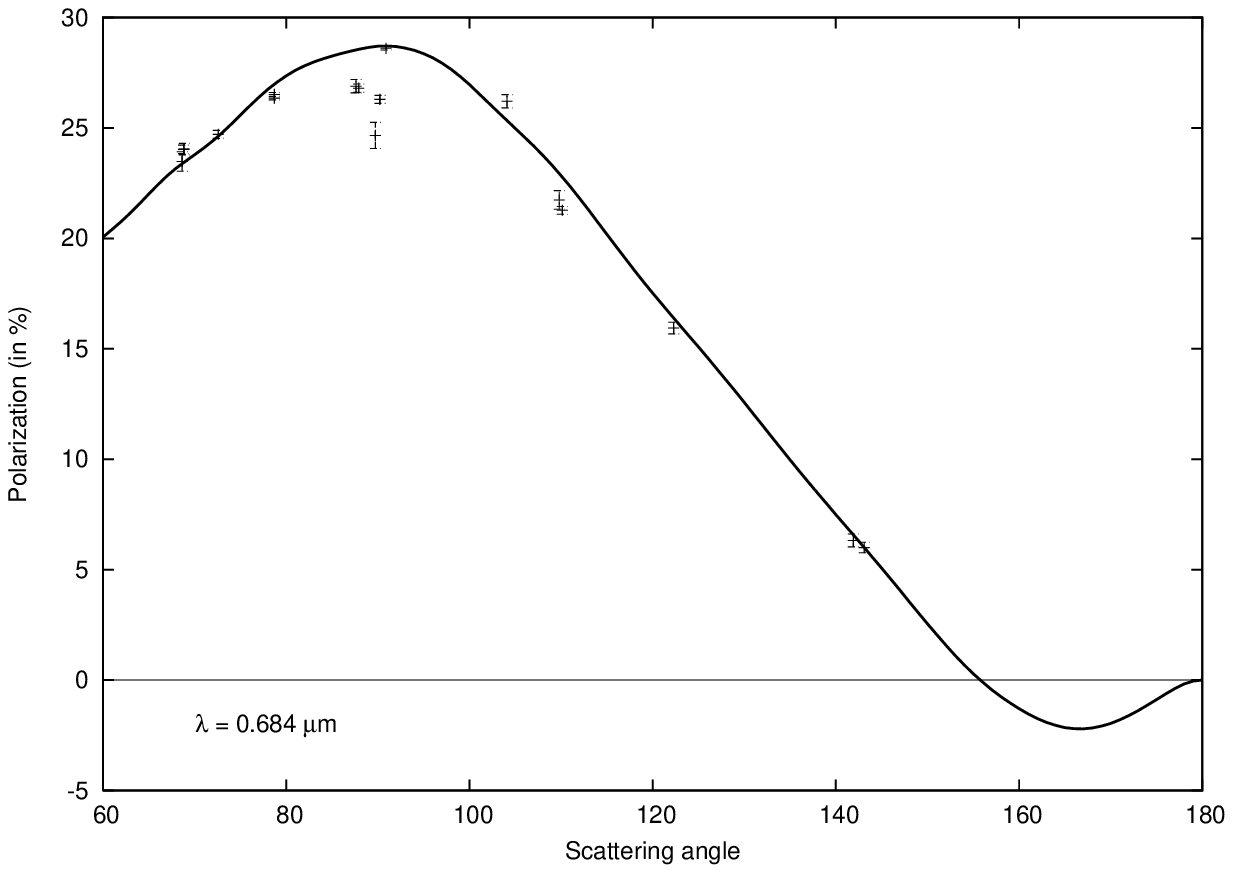}
   \caption{Polarization values as observed at wavelength $\lambda$ = 0.684$\mu m$ for
comet Hyakutake by Joshi et al. (1997), Kiselev \& Velichko (1998) and Manset \& Bastien (2000). The solid  curve represents
 the best-fitting polarization curve
 obtained  for BCCA particles
 with 128 monomers  for a size distribution $n(r) \sim r^{-3}$ for $0.88\mu m \le r \le 0.94 \mu m$ at
 $\lambda = 0.684$ $\mu m$. Here $n$ = 1.695, $k$ = 0.100.}\label{figexample}
    \end{figure*}



\section{Discussion:}
\label{sect:discussion}
The negative polarization behaviour of comet is one of the major
features observed in comets. Several comets show negative polarization
beyond 157$^0$ scattering angle (Kikuchi et al., 1987; Chernova
et al., 1993; Ganesh et al., 1998 etc.).  Interestingly, all comets show very similar
characteristics of negative polarization (minimum value of polarization $\sim$ -- 2 $\%$ near 170$^0$ and inversion angle at 20 -- 22$^0$). Comet Hyakutake was observed
over a wide scattering angle range (68.6$^0$ -- 143.1$^0$), but there was no observation recorded
beyond 143.1$^0$ (Joshi et al., 1997; Kiselev \& Velichko 1998 and Manset \& Bastien 2000).
In the present work, it is interesting to observe that the used dust aggregate model
reproduce the negative polarization behaviour beyond 157$^0$.

The strength of the silicate feature  is defined as the ratio of
the flux between 10 and 11 $\mu m$ to that of the underlying
continuum (Lisse 2002; Sitko et al., 2004; Kolokolova et al., 2007).
The silicate feature strength of  Comet Hyakutake  is $> 1.5$ (Lisse 2002)
whereas the values for Comet Levy 1990XX and Comet Hale-Bopp are given
by 1.8 (Harker et al., 1999) and 2.16 (Sitko et al., 2004). Comet
Hale-Bopp  is an intrinsically  bright comet, with polarization
values much higher than those of other comets. It has been found
that Comet Hale-Bopp shows the highest silicate feature strength.
The strong silicate feature indicates high abundance of silicates
in the dust. It can be seen  that the refractive
indices coming out from the present calculation is closed to the refractive indices of silicates and organics.
Again the \textit{in situ} measurements of comet Halley (Lamy et al. 1987) and the `Stardust'
returned samples  of comet Wild 2 (Zolensky et al., 2006) showed the presence of a mixture of
silicates and organic refractory in cometary dust. Thus, our model calculations represent the more
realistic type of grains which may be considered as a mixture of silicates and carbonaceous materials.
It is to be noted that the presence of negative polarization in the backscatter domain has been
commonly attributed to silicates or dirty ice grains (Kimura et al., 2006).

It has been investigated that the  aggregate dust model can well fit the observed polarization data of comet Hyakutake when the size parameter of the monomer, $x$ $\sim$ 1.56 -- 1.70. Thus the size ranges of the monomer differ for three wavelengths which is  unlikely.  The proposed model can be further developed if we take a mixture of compact spheroidal grains and aggregates over a wide size range which Lasue et al. (2009) used in their paper. They studied comet Halley and comet Hale-Bopp using a mixture of fluffy
aggregates and compact solid grains and successfully explained the observed polarization characteristics of two comets. In a follow-up paper, we also plan to model cometary dust as a mixture of aggregates and compact particles.

\section{Conclusions}
\label{sect:conclusion}
\begin{enumerate}
  \item  The size parameter of the monomer,  $x$ $\sim$ 1.56 -- 1.70, turned out to be most
suitable which provides the best  fits to the observed polarization data of comet Hyakutake at
three wavelengths $\lambda = 0.365$$\mu m$, 0.485$\mu m$ and 0.684$\mu m$. This correspond to  $0.090\mu m \le a_m \le 0.098 \mu m$ at $\lambda = 0.365$$\mu m$; $0.120\mu m \le a_m \le 0.131 \mu m$ at $\lambda = 0.485$$\mu m$ and
$0.174\mu m \le a_m \le 0.186 \mu m$ at $\lambda = 0.684$$\mu m$.

  \item  The best fit refractive indices
coming out from the present analysis are $n = 1.745$ and $k =
0.095$ for N = 128 at $\lambda = 0.365$ $\mu m$; $n = 1.743$ and $k =
0.100$ for N = 128 at $\lambda = 0.485$ $\mu m$ and $n = 1.695$
and $k = 0.100$ for N = 128 at $\lambda = 0.684$ $\mu m$. These
values resemble the  mixture of silicates and carbonaceous compounds.

\item  The negative polarization values have been successfully generated for $\theta > 157^0$ at three wavelengths.

\item We plan a follow-up paper where computations will be made considering a mixture of  aggregates and compact spheroidal  particles over a wide size range of the partcles.

\end{enumerate}

\section*{Acknowledgments}
The authors HSD and AKS  acknowledge  Inter University Centre for
Astronomy and Astrophysics (IUCAA), Pune for its associateship
programme.   The authors acknowledge T. Mukai and Y. Okada for
help on the execution of BPCA and BCCA codes. The authors are
thankful to D. Mackowski, K. Fuller, and M. Mishchenko, who made
their superposition T-matrix code publicly available. The authors highly acknowledge the referee of the paper
for his valuable suggestions and comments for which the quality of the paper has been improved.

\label{lastpage}

\end{document}